\documentclass[preprint,showpacs,preprintnumbers,amsmath,amssymb,showkeys,nofootinbib]{revtex4}
   
\usepackage{graphicx}% Include figure files 
\usepackage{dcolumn}% Align table columns on decimal point
\usepackage{bm}% bold math
\usepackage{braket}
 
\begin{document} 

\noindent

\preprint{}

\title{Estimation independence as an axiom for quantum uncertainty}% Force line breaks with \\

\author{Agung Budiyono} 
\email{agungbymlati@gmail.com}
\affiliation{Research Center for Nanoscience and Nanotechnology, Bandung Institute of Technology, Bandung, 40132 Indonesia; \\
Edelstein Center, Hebrew University of Jerusalem, Jerusalem, 91904 Israel;\\
Department of Engineering Physics, Bandung Institute of Technology, Bandung, 40132 Indonesia;\\
and Kubus Computing and Research, Juwana, Pati, 59185 Indonesia}
 
\date{\today}% It is always \today, today, 
   
\begin{abstract}   

Quantum uncertainty is the cornerstone of quantum mechanics which underlies many counterintuitive nonclassical phenomena. Recent studies remarkably showed that it also fundamentally limits nonclassical correlation, and crucially, a deviation from its exact form may lead to a violation of the second law of thermodynamics. Are there deep and natural principles which uniquely determines its form? Here we work within a general epistemic framework for a class of nonclassical theories, introducing an epistemic restriction to an otherwise classical theory, so that the distributions of positions are irreducibly parameterized by the underlying momentum fields. It was recently shown that the mathematics of quantum mechanics formally arises within an operational scheme, wherein an agent makes a specific estimation of the momentum given information on the positions and the experimental settings. Moreover, quantum uncertainty can be traced back to the `specific' choice of estimator and the associated estimation error. In the present work, we show that a plausible principle of estimation independence, which requires that the estimation of momentum of one system must be independent of the position of another system independently prepared of the first, singles out the specific forms of the estimator, and especially the estimation error up to its strength given by a global-nonseparable random variable on the order of Planck constant. 
     
\end{abstract}     
 
%Use showkeys class option if keyword
\maketitle  

\section{Introduction}  
 
Quantum uncertainty is one of the distinctive features of quantum mechanics which radically sets it apart from classical mechanics \cite{Heisenberg UR,Kennard UR}. Formally, it is the result of the non-commutative structure of quantum observables acting on complex Hilbert space, which rules out the preparation of a quantum state with all observables having sharp values. Deeper understanding of its physical and/or informational origin and meaning holds the key to better understand classically puzzling microscopic phenomena \cite{Ahoronov-Daniel book}, and its harnessing to emerging quantum technologies \cite{BB84,Berta UR with memory,Coles et al review}. Quantum uncertainty is argued to underlie nonclassical correlation \cite{Tsirelson noncommutativity and violation of Bell inequality,Seevinck noncommutativity-correlation trade-off,Fine-Bell theorem,Cavalcanti non-commutativity and Bell inequality} and to enable nonclassical dynamics \cite{Dahlsten et al paper}. But, at the same time, it has been suggested that quantum uncertainty also constrains the strength of the nonclassical correlation \cite{Oppenheim-Wehner entropic UR and QS,Ver Steeg relaxing uncertainty relation}, and indirectly uniquely determines the form of quantum dynamics \cite{Agung-Daniel model,Hall-Reginatto model}. Moreover, crucially, it was shown recently that a deviation from the exact form of quantum uncertainty may strikingly allow for the creation of thermodynamics cycle with a positive work gain, violating the second law of thermodynamics \cite{Hanggi UR second law}. This begs a question: are there deep, profound, and natural principles determining its exact form? 

Clarifying the above foundational question is clearly very important to advance in the long-running attempts to search for a set of transparent, reasonable, and compelling physical axioms which uniquely define quantum mechanics \cite{Wheeler: why,Popescu-Rohrlich axioms,Rovelli relational QM,Zeilinger axioms,Caves-Fuchs axioms,Hardy axioms,Fuchs axioms,Spekkens toy model with epistemic restriction,Barrett axioms,Dakic-Brukner axioms,Paterek axioms,Chiribella axioms,Masanes axioms,de la Torre axioms,Bartlett reconstruction of Gaussian QM with epistemic restriction,Reginatto estimation scheme,Hall-Reginatto model,Markopolou-Smolin quantum from cosmos,Smolin quantum from cosmos,Nelson stochastic mechanics,Frieden model, deRaedt model,Agung-Daniel model,Goyal information geometry,Caticha quantum from MEP,Hohn quantum from question,Selby quantum from diagram}, rather than the infamously obscure mathematical axioms of the standard complex Hilbert space formalism. This reconstruction research program is pivotal to better understand the meaning of the abstract rules of quantum mechanics, and receives a strong renewed interest in the past decades owing to the astonishing progress in quantum information theory. It was shown remarkably that: introducing a nonlinearity in the Schr\"odinger equation may imply superluminal signalling \cite{Gisin nonlinearity - signaling,Polchinski nonlinearity - signaling,Czachor nonlinearity - signaling,Simon no-signaling imply linearity,Mielnik nonlinearity - signaling} and also a violation of the second law of thermodynamics \cite{Peres nonlinearity violates 2nd law}; nonlinearity or a deviation from Born's quadratic law may lead to computational schemes fundamentally much faster than quantum computation \cite{Abrams-Lloyd nonlinearity - fast computation,Aaronson nonlinearity-nonunitary - fast computation}; introducing a fundamental nonHermiticity in quantum Hamiltonian may also contradict non-signaling \cite{Lee nonHermitian - signaling}; and, allowing no-signalling stronger than quantum correlation \cite{Popescu-Rohrlich axioms} may imply implausible computational power \cite{Popescu review,Dam informational approach Tsirelson bound,Brassard informational approach Tsirelson bound,Buhrman superstrong cryptography,Linden nonlocal computation,Brunner trivial communication,Pawlowski informational approach Tsirelson bound,Gross trivial dynamics with superstrong correlation}. These observations have led to the belief that quantum theory might be an `island' in the space of physical theories \cite{Aaronson nonlinearity-nonunitary - fast computation,Aaronson QM is an island}, isolated by some transparent physical axioms. 

In the present work, we study the above problem within a general epistemic framework for a class of nonclassical theories, introducing an epistemic restriction to an otherwise classical theory, so that the allowed forms of distribution of positions are irreducibly parameterized by the underlying momentum fields \cite{Agung-Daniel model}. We showed in the previous work \cite{Agung epistemic interpretation} that the abstract mathematical rules of quantum mechanics formally emerges within an operational scheme wherein an agent makes an estimation of the underlying momentum field, given information on the conjugate positions, with `specific' estimator and estimation error (given respectively in Eqs. (\ref{weakly unbiased best estimator}) and (\ref{estimation error})). Moreover, quantum uncertainty relation is shown to originate from the irreducible trade-off between the mean squared (MS) errors of simultaneous estimations of momentum field and mean position, which is deduced directly from the alluded above specific estimator and estimation error. 

Within the above epistemic or informational reconstruction of quantum mechanics, the search for transparent physical principles underlying quantum uncertainty therefore translates into: why that `specific' forms of estimator and the estimation error for the estimation of momentum given positions postulated in Ref. \cite{Agung epistemic interpretation}? Are there deep and natural principles which rule out any other alternative choices? Here we argue that a transparent and plausible principle of estimation independence, which requires that independent preparations must entail independent estimations, singles out the specific forms of the estimator, and especially the estimation error up to its strength given by a global-nonseparable random variable on the order of Planck constant. A violation of the principle of estimation independence may imply an unnatural information gain: namely, an agent may benefit from the information on the position of a system, to reduce the error of estimating the momentum of another system, prepared independently of the first. A simple example on an estimation scheme violating the principle of the estimation independence implying a non-trivial deviation from quantum uncertainty relation is given.       

\section{Quantum uncertainty as an irreducible limitation on simultaneous estimations under epistemic restriction\label{Quantum uncertainty as a fundamental limitation on simultaneous estimations under epistemic restriction}}

First, consider a classical system of $N$ degrees of freedom with a spatial configuration $q=(q_1,\dots,q_N)$ and the conjugate momentum  $p=(p_1,\dots,p_N)$. Within the Hamilton-Jacobi formalism of classical mechanics  \cite{Rund book: Hamilton-Jacobi formalism}, the momentum field can be expressed as   
\begin{eqnarray}
\tilde{p}_{\rm C}(q,t)=\partial_qS_{\rm C}(q,t), 
\label{HJ condition} 
\end{eqnarray}
where $\tilde{p}_{\rm C}=(\tilde{p}_{{\rm C}_1},\dots,\tilde{p}_{{\rm C}_N})$, $\partial_q=(\partial_{q_1},\dots,\partial_{q_N})$, and $S_{\rm C}(q,t)$ is the Hamilton's principal function. (To avoid confusion, we use $\tilde{p}$ to denote the functional form of momentum fields, whereas $p$ denotes a specific value of momentum.) Moreover, for a system with a classical Hamiltonian $H(q,p)$, the time evolution of $S_{\rm C}(q,t)$ is governed by the Hamilton-Jacobi equation: $-\partial_tS_{\rm C}(q,t)=H(q,p)=H(q,\partial_qS_{\rm C}(q,t))$. Solving the Hamilton-Jacobi equation for $S_{\rm C}(q,t)$, a single trajectory in configuration space is singled out by selecting the positions $q=q_0$ at an arbitrary time $t=t_0$, and integrating Eq. (\ref{HJ condition}). The Hamilton-Jacobi formalism of classical mechanics thus describes an ensemble of trajectories, obtained by varying $q=q_0$ at $t=t_0$ in an ensemble of identical preprations, all following a momentum field $\tilde{p}_{\rm C}(q,t)$ characterized by a single Hamilton's principal function $S_{\rm C}(q,t)$. 

Next, consider such an ensemble of trajectories, and denote the probability distribution of the positions $q$ at time $t$ as $\rho(q,t)$. In classical mechanics, given a momentum field $\tilde{p}_{\rm C}(q)$ (here and below, trivial dependence on time is notationally omitted), it is clear from Eq. (\ref{HJ condition}) that, in principle, an agent can prepare an ensemble of trajectories with an arbitrary distribution of positions $\rho(q)$. Namely, each trajectory in the momentum field $\tilde{p}_{\rm C}(q)$ can be assigned an arbitrary weight $\rho(q)$. Hence, in classical mechanics, the probability distribution of positions $\rho(q)$ is in principle independent of, thus is not fundamentally parameterized by, the underlying momentum field $\tilde{p}_{\rm C}(q)$. We call such a fundamental freedom to choose an arbitrary distribution of positions independent of the underlying momentum field as ``epistemic freedom'', and argue that it is a basic principle (implicitly assumed) in classical mechanics \cite{Agung-Daniel model}.   

Partly inspired by recent remarkable results that a significant fraction of microscopic phenomena traditionally regarded as specifically quantum can in fact be explained from classical statistical models with some epistemic restrictions \cite{Hardy classical reconstruction of teleportation,Emerson epistemic restriction,Spekkens toy model with epistemic restriction,Bartlett reconstruction of Gaussian QM with epistemic restriction}, it is shown in Ref. \cite{Agung-Daniel model} that the mathematical formalism of spinless nonrelativistic quantum mechanics can be obtained by abandoning the above principle of epistemic freedom. First, we assume that in microscopic world, the momentum field $\tilde{p}(q,t;\xi)$ fluctuates randomly induced by a global-nonseparable ontic random variable $\xi$. We then assume that the ensemble of trajectories following the momentum field no longer respects the principle of epistemic freedom, but experiences a fundamental `epistemic restriction' \cite{Agung-Daniel model}: namely, given a momentum field, unlike in classical mechanics, it is in principle no longer possible to assign each trajectory in the momentum field an arbitrary weight. The allowed distributions of positions that an agent can prepare, therefore fundamentally depend on, thus are irreducibly parameterized by, the underlying momentum fields $\tilde{p}(q;\xi)$. To make explicit this intrinsic dependence via restriction, we write the probability distribution of positions at time $t$ as 
\begin{eqnarray}
\rho_{\tilde{p}}(q,t), \nonumber
\label{general epistemic restriction}
\end{eqnarray}
with a subscript $\tilde{p}$. We further assume that in the formal limit of vanishing global fluctuation $\xi$, we have $\lim_{\xi\rightarrow 0}\tilde{p}(q;\xi)=\tilde{p}_{\rm C}(q)=\partial_qS_{\rm C}$ of Eq. (\ref{HJ condition}), and $\lim_{\xi\rightarrow 0}\rho_{\tilde{p}}(q)=\rho(q)$, so that we regain classical mechanics with the epistemic freedom. 

Now, suppose an agent has an access to the value of $q$. Since $q$ is sampled from $\rho_{\tilde{p}}(q)$, it must somehow carry some information about the conjugate momentum field $\tilde{p}(q;\xi)$ parameterizing $\rho_{\tilde{p}}(q)$. How can the agent use her information on $q$, in the most reasonable way, to estimate $\tilde{p}(q;\xi)$ resulting in the preparation? This is just a parameter estimation problem in the (classical) statistical-information theory \cite{Papoulis and Pillai book on probability and statistics}. Within this operational scheme of estimation under epistemic restriction, we showed in Ref. \cite{Agung epistemic interpretation} that the abstract rules of quantum mechanics in complex Hilbert space, including quantum uncertainty, formally arises, as the agent makes a specific estimation of the momentum given information on the conjugate positions, minimizing the MS error, to be summarized below for later purpose.   

First, to have a consistent classical limit, noting the fact that in macroscopic regime the momentum field takes the form of Eq. (\ref{HJ condition}), we assume that given $q$, the estimator for $\tilde{p}(q;\xi)$ has the form 
\begin{eqnarray}
\overline{p}(q,t)\doteq\partial_qS(q,t), 
\label{weakly unbiased best estimator}
\end{eqnarray}
where $\overline{p}=(\overline{p}_1,\dots,\overline{p}_N)$, and $S(q,t)$ is a real-valued scalar function of $(q,t)$, so that in the classical limit the estimator is expected to approach the gradient of the Hamilton's principal function, i.e., $\overline{p}(q)=\partial_qS (q)\rightarrow \partial_qS_{\rm C}(q)$, recovering Eq. (\ref{HJ condition}). Furthermore, given $q$, the estimation error in a single-shot estimate of $\tilde{p}(q;\xi)$ with the estimator $\overline{p}(q)\doteq\partial_qS(q)$ is postulated to have the following form \cite{Agung epistemic interpretation}:
\begin{eqnarray}
\epsilon_p(q;\xi)\doteq \tilde{p}(q;\xi)-\partial_qS(q)=\frac{\xi}{2}\partial_q\ln\rho_{\tilde{p}}(q).
\label{estimation error}
\end{eqnarray}
Let us assume that $\xi$ fluctuates randomly on a microscopic timescale, with a probability distribution $\chi(\xi)$, so that its first and second moments are independent of $(q,t)$, given by \cite{Agung-Daniel model} 
\begin{equation}
\overline{\xi}\doteq\int {\rm d}\xi~\xi~\chi(\xi)=0,~~~\overline{\xi^2}=\hbar^2.
\label{Planck constant}
\end{equation}
In the regime when the estimation error is much smaller than the estimator, i.e., $|\partial_qS(q)|\gg|\frac{\xi}{2}\partial_q\ln\rho_{\tilde{p}}(q)|$, one effectively regains the classical relation of Eq. (\ref{HJ condition}): $\tilde{p}(q)\approx\overline{p}(q)=\partial_qS(q)$. Assuming further that $\rho_{\tilde{p}}(q)$ is vanishing at the boundary, one can see that the above estimation error satisfies a desirable weak unbiased condition, i.e., the average of the estimation error over $\rho_{\tilde{p}}(q)$ for any $\xi$ is vanishing: $\int{\rm d}q\epsilon_p(q;\xi)\rho_{\tilde{p}}(q)=\frac{\xi}{2}\int{\rm d}q\partial_q\rho_{\tilde{p}}(q)=0$, ${\rm d}q\doteq{\rm d}q_1{\rm d}q_2\cdots{\rm d}q_N$. Finally, one can also show that the estimator indeed minimizes the associated MS error defined as $\mathcal{E}_{p_j}^2\doteq\int{\rm d}q{\rm d}\xi\epsilon_{p_j}(q;\xi)^2\chi(\xi)\rho_{\tilde{p}}(q)$ \cite{Agung epistemic interpretation}. 

We have argued in Ref. \cite{Agung epistemic interpretation} that the above specific operational scheme of estimation of momentum given information on the positions, combined with the principle of conservation of average energy and trajectories suggested by the Bayesian reasoning given the experimental settings, lead to the epistemic reconstruction of the abstract mathematical rules of quantum mechanics. We refer to Refs. \cite{Agung-Daniel model,Agung epistemic interpretation} for the detailed mathematical derivations. For later purpose, we only need to put forward an important result that, within the epistemic reconstruction based on the operational scheme of estimation under epistemic restriction, the quantum wave function $\psi(q,t)$ characterizing a preparation is mathematically reconstructed from the estimator of Eq. (\ref{weakly unbiased best estimator}) and the estimation error of Eq. (\ref{estimation error}), via $\big(S(q,t),\rho_{\tilde{p}}(q,t)\big)$, as 
\begin{equation} 
\psi(q,t)\doteq\sqrt{\rho_{\tilde{p}}(q,t)}\exp(iS(q,t)/\hbar). 
\label{wave function}
\end{equation}
It is therefore clear that, by construction, quantum wave function is not an agent-independent objective property of the system, but a mathematical tool which represents the agent's weakly unbiased best estimation of the momentum field arising in her preparation, based on information on the conjugate positions, under epistemic restriction \cite{Agung epistemic interpretation}. Moreover, from Eq. (\ref{wave function}), the Born's quadratic law, i.e., $\rho_{\tilde{p}}(q)=|\psi(q)|^2$, is also valid by construction. 

Furthermore, within the epistemic reconstruction, quantum uncertainty can be traced back to the specific forms of estimator of Eq. (\ref{weakly unbiased best estimator}) and estimation error of Eq. (\ref{estimation error}), as follows. First, from Eqs. (\ref{estimation error}) and (\ref{Planck constant}), the MS error of the estimation of the momentum field $\tilde{p}_j(q;\xi)$ with the estimator $\partial_{q_j}S(q)$, $j=1,\dots,N$, can be computed to get
\begin{eqnarray}
\mathcal{E}_{p_j}^2=\frac{\hbar^2}{4}J_{q_j}, 
\label{information trade-off}
\end{eqnarray}
where $J_{q_j}$ is the Fisher information about the mean position $q_{o_j}\doteq\int{\rm d}q q_{o_j}\rho_{\tilde{p}}(q)$ parameterizing $\rho_{\tilde{p}}(q)$ defined as \cite{Papoulis and Pillai book on probability and statistics} 
\begin{eqnarray}
J_{q_j}\doteq\int{\rm d}q\big(\partial_{q_j}\ln\rho_{\tilde{p}}(q)\big)^2\rho_{\tilde{p}}(q). 
\label{Fisher information about the mean position}
\end{eqnarray}
Remarkably, as argued in Ref. \cite{Agung ERPS distribution}, the estimator $\partial_{q_j}S(q)$ of the momentum field, and the associated MS error $\mathcal{E}_{p_j}^2$, can be operationally probed in experiment via weak momentum value measurement \cite{Aharonov weak value,Lundeen complex weak value,Jozsa complex weak value}. Eq. (\ref{information trade-off}) transparently describes a kind of information trade-off on the order of Planck constant \cite{Agung epistemic interpretation}. Namely, the smaller (larger) $\mathcal{E}_p^2$, i.e., the sharper (poorer) the agent's estimate of the momentum, the smaller (larger) $J_q$, i.e., the poorer (better) her knowledge about the position; it thus somehow already quantifies Bohr's idea on the complementarity between momentum and position. On the other hand, in an estimation of mean position $q_{o_j}$ with the unbiased estimator $q_j$, the associated MS error must satisfy the Cram\'er-Rao inequality: $\mathcal{E}_{q_j}^2\doteq\int{\rm d}q(q_j-q_{o_j})^2\rho_{\tilde{p}}(q)\ge 1/J_{q_j}$, $j=1,\dots,N$. Combining this with Eq. (\ref{information trade-off}), one therefore obtains 
\begin{eqnarray}
\mathcal{E}_{p_j}^2\mathcal{E}_{q_j}^2\ge \frac{\hbar^2}{4},
\label{trade off between MS error of position and momentum}
\end{eqnarray}
$j=1,\dots,N$, describing the impossibility for an agent to have sharp simultaneous estimations of momentum field and mean position

Next, let us denote the conventional ensemble average of a physical quantity $O(p,q)$ in the statistical model as $\braket{O}_{\{S,\rho_{\tilde{p}}\}}\doteq\int{\rm d}q{\rm d}\xi{\rm d}pO(p,q){\rm P}_{\{S,\rho_{\tilde{p}}\}}(p,q|\xi)\chi(\xi)$, where ${\rm P}_{\{S,\rho_{\tilde{p}}\}}(p,q|\xi)$ is the epistemically restricted phase space distribution defined as, noting Eq. (\ref{estimation error}), ${\rm P}_{\{S,\rho_{\tilde{p}}\}}(p,q|\xi)\doteq\prod_{j=1}^N\delta\big(p_j-\partial_{q_j}S-\frac{\xi}{2}\partial_{q_j}\ln\rho_{\tilde{p}}\big)\rho_{\tilde{p}}(q)$ \cite{Agung ERPS distribution}. Then, for $j=1,\dots,N$, using Eq. (\ref{wave function}), it is straightforward to show the following mathematical identities: 
\begin{eqnarray}
\sigma_{q_j}^2&=&\sigma_{\hat{q}_j}^2=\mathcal{E}_{q_j}^2,\nonumber\\
\sigma_{p_j}^2&=&\sigma_{\hat{p}_j}^2=\mathcal{E}_{p_j}^2+\Delta_{p_j}^2,
\label{measurement as unbiased estimates}
\end{eqnarray}
where $\sigma_{q_j}^2\doteq\braket{(q_j-\braket{q_j}_{\{S,\rho_{\tilde{p}}\}})^2}_{\{S,\rho_{\tilde{p}}\}}$, $\sigma_{p_j}^2\doteq\braket{(p_j-\braket{p_j}_{\{S,\rho_{\tilde{p}}\}})^2}_{\{S,\rho_{\tilde{p}}\}}$ are the variances of position and momentum over the epistemically restricted ensemble of trajectories, $\sigma_{\hat{q}_j}^2\doteq\braket{\psi|(\hat{q}_j-\braket{\psi|\hat{q}_j|\psi})^2|\psi}$, $\sigma_{\hat{p}_j}^2\doteq\braket{\psi|(\hat{p}_j-\braket{\psi|\hat{p}_j|\psi})^2|\psi}$ are the associated quantum variances, and $\Delta_{p_j}^2\doteq\int{\rm d}q\big(\partial_{q_j}S(q)-\int{\rm d}q'\partial_{q'_j}S(q')\rho_{\tilde{p}}(q')\big)^2\rho_{\tilde{p}}(q)\ge 0$ is just the dispersion of the estimator $\overline{p}_j(q)=\partial_{q_j}S(q)$ \cite{Agung epistemic interpretation}. 

Finally, combining the two equations in Eq. (\ref{measurement as unbiased estimates}), one obtains, by the virtue of Eq. (\ref{trade off between MS error of position and momentum}), the Heinseberg-Kennard uncertainty relation 
\begin{eqnarray}
\sigma_{p_j}^2\sigma_{q_j}^2&=&\sigma_{\hat{p}_j}^2\sigma_{\hat{q}_j}^2=\mathcal{E}_{p_j}^2\mathcal{E}_{q_j}^2+\Delta_{p_j}^2\mathcal{E}_{q_j}^2\nonumber\\
&\ge&\hbar^2/4+\Delta_{p_j}^2\mathcal{E}_{q_j}^2\ge\hbar^2/4,
\label{HK UR}
\end{eqnarray} 
$j=1,\dots,N$. A derivation of the Schr\"odinger-Robertson uncertainty relation between the momentum and position within the above estimation scheme is given in Appendix \ref{Schroedinger-Robertson uncertainty relation}. Hence, the results show that quantum uncertainty is deeply related with the Cram\'er-Rao inequality limiting the agent's estimation about her preparation. One can also show that Gaussian wave function represents the ``efficient'' simultaneous estimation of momentum field and mean position, achieving the Cram\'er-Rao bounds of the associated MS errors \cite{Agung epistemic interpretation}. We note that a derivation of quantum uncertainty from Cram\'er-Rao inequality is also reported in Ref. \cite{Frowis UR from CR inequality}. However, unlike our derivation which does not presume any quantum structures, the authors in Ref. \cite{Frowis UR from CR inequality} assumes quantum trace rule, and unitary quantum dynamics.   

Let us compare the above epistemic reconstruction and interpretation of quantum mechanics based on the operational scheme of estimation under epistemic restriction, with the most well-known realist-causal de Broglie-Bohm interpretation \cite{Bohmian mechanics}, here on referred to as  Bohmian mechanics. Bohmian mechanics starts by taking the Schr\"odinger equation for granted, and assumes that the wave function $\psi(q,t)$ is an objective physical-real field (living in multi-dimensional configuration space), whose dynamics physically-causally guides the time evolution of the positions of the particles as 
\begin{eqnarray}
\tilde{p}_{\rm BM}(q,t)=\hbar\partial_q{\rm Arg}\{\psi(q,t)\}=\partial_qS(q,t),
\label{Bohmian guidance relation}
\end{eqnarray} 
where $\tilde{p}_{\rm BM}(q,t)$ is the objective momentum of the particles independent of measurement at time $t$ with positions $q$, and we have used Eq. (\ref{wave function}) in the second equality (cf. Eq. (\ref{weakly unbiased best estimator})). Moreover, Bohmian mechanics also postulates the Born's quadratic law ${\rm P}_{\psi}(q)=|\psi(q)|^2$ (to be valid at least at some initial time). Note that while there are attempts to justify the above guidance relation \cite{Duerr why Bohmian velocity,Wiseman why Bohmian velocity}, it is not entirely clear why it has to be of the specific form given by Eq. (\ref{Bohmian guidance relation}). In fact, there are infinitely many different versions of Bohmian mechanics which reproduce the prediction of quantum mechanics \cite{Deotto alternative to Bohmian guidance relation}. Moreover, there is a question why the distribution of the positions must be given by the Born's quadratic law \cite{Valentini Born's quadratic law,Duerr Born's quadratic law}. 

In sharp contrast to Bohmian mechanics, within the epistemic interpretation based the operational scheme of estimation under epistemic restriction discussed in the present paper, the Schr\"odinger equation is derived rather than postulated as a fundamental law; moreover, it does not describe a real-physical processes, but a normative rule to rationally update the agent's estimation when she does not make any measurement \cite{Agung epistemic interpretation}. Hence, unlike in Bohmian mechanics, by construction, the wave function defined in Eq. (\ref{wave function}) is not an agent-independent real-physical thing, but a mathematical object which summarizes the agent's estimation about her preparation; it lives in the agent's mind rather than in the physical space \cite{Agung epistemic interpretation}. Moreover, Eq. (\ref{weakly unbiased best estimator}) is not a  physical guidance relation, but a prescription to guide the agent's belief: namely, $\partial_qS(q)$ is not the objective momentum of the particles at $q$ as favoured by Bohmian mechanics, but it is the agent's best estimate of the momentum of the particle given $q$. Furthermore, we have assumed a global-nonseparable random variable $\xi$ on the order of the Planck constant which is missing in the standard Bohmian mechanics; it plays a central role in our epistemic reconstruction based on the operational scheme of estimation, giving the strength of the error of single-shot estimation of momentum as in Eq. (\ref{estimation error}), and the statistical origin of the lower bound of the product between the MS errors of simultaneous estimation of momentum field and mean position of Eq. (\ref{trade off between MS error of position and momentum}) implying the Heisenberg-Kennard uncertainty relation of Eq. (\ref{HK UR}). Within the epistemic reconstruction, the agent's best estimate of momentum therefore coincides with the Bohmian momentum, with an error on the order of Planck constant. Further important conceptual difference between our epistemic reconstruction-interpretation and the de Broglie-Bohm realist-causal interpretation is suggested at the last paragraph of the present article.  

\section{Independent preparations and principle of estimation independence  \label{Principle of estimation independence}}     

The above epistemic reconstruction within the operational scheme of estimation under epistemic restriction shows that quantum uncertainty, the cornerstone of quantum mechanics, originates from the irreducible trade-off between the MS errors of simultaneous estimations of momentum field and mean position of Eq. (\ref{trade off between MS error of position and momentum}). Moreover, crucially, the latter is straightforwardly obtained from the specific form of estimator of Eq. (\ref{weakly unbiased best estimator}), and especially the apparently ad-hoc specific form of estimation error of Eq. (\ref{estimation error}), for the estimation of momentum given information on the conjugate positions. We show below, as a concrete example, that a different choice of estimation error indeed leads to a different form of uncertainty relation. Our main interest in the present work is then: why, to obtain the exact form of quantum uncertainty | a deviation from which may strikingly imply a violation of the second law of thermodynamics \cite{Hanggi UR second law} | the estimator and especially the estimation error for the estimation of momentum given the positions, must take that `specific' forms? Are there deep and natural principles which rule out any alternative choices? In this section we argue for a positive answer. 

To study this foundational question, we consider two systems, referred to as system 1 and system 2, with a spatial configuration $q=(q_1,q_2)$, prepared independently of each other. Let us first discuss the classical mechanics description of such pairs of independent preparations. We recall that in the Hamilton-Jacobi formalism of classical mechanics, the Hamilton's principal function $S_{\rm C}(q;t)$ plays a central role, describing an ensemble of trajectories arising in repeated identical preparations. It is therefore desirable to express the independent preparations in terms of the Hamilton's principal function. To this end, since independent preparations of two non-interacting systems in classical mechanics is described by a decomposable Lagrangian, i.e., $L(q_1,q_2,\dot{q}_1,\dot{q}_2)=L_1(q_1,\dot{q}_1)+L_2(q_2,\dot{q}_2)$, where $\dot{q}\doteq{\rm d}q/{\rm d}t$, the associated Hamilton's principal function must also be decomposable, i.e., $S_{\rm C}(q_1,q_2,t)=\int^{(q_1,t)}{\rm d}t'L_1(q'_1,\dot{q}'_1)+\int^{(q_2,t)}{\rm d}t'L_2(q'_2,\dot{q}'_2)=S_{\rm C}(q_1,t)+S_{\rm C}(q_2,t)$. Moreover, in classical mechanics, the distribution of positions obtained in an ensemble of such independent preparations is naturally separable, i.e., $\rho(q_1,q_2)=\rho_1(q_1)\rho_2(q_2)$. Hence, in the Hamilton-Jacobi formalism of classical mechanics, the two fundamental entities describing the ensemble of trajectories arising in an ensemble of independent preparations, namely $S_{\rm C}(q)$ and $\rho(q)$, are respectively decomposable and separable. 

Now, we upgrade the above intuitive basic features of classical mechanics to apply also in microscopic world. Namely, first, we assume that there is a scalar function $S(q)$, replacing the role of $S_{\rm C}(q)$, so that for two systems prepared independently of each other, it is decomposable into that of each system, i.e.,
\begin{eqnarray}
S(q_1,q_2)=S_1(q_1)+S_2(q_2). 
\label{phase decomposability for independent preparations}
\end{eqnarray}
Second, we also assume that in microscopic world, the probability distribution of positions arising in independent preparations of two systems is separable as in classical mechanics, i.e.,
\begin{eqnarray}
\rho_{\tilde{p}}(q_1,q_2)=\rho_{\tilde{p}_1}(q_1)\rho_{\tilde{p}_2}(q_2).
\label{separable amplitude for independent preparation}
\end{eqnarray}
We admit that the above two assumptions are heruistic, motivated intuitively by our desire to have a conceptually and formally smooth quantum-classical correspondence and transition. If we define the wave function as in Eq. (\ref{wave function}), the above two assumptions amount to the assumption that the wave function associated with independent preparations of two systems is factorizable (unentangled), i.e., $\psi(q_1,q_2)=\sqrt{\rho_{\tilde{p}}}e^{\frac{i}{\hbar}S}=\sqrt{\rho_{\tilde{p}_1}\rho_{\tilde{p}_2}}e^{\frac{i}{\hbar}(S_1+S_2)}=\psi_1(q_1)\psi_2(q_2)$. 
 
Keeping the above assumptions in mind, let us proceed to define the principle of estimation independence. Consider again the independent preparations of two systems. Then, within the operational framework of parameter estimation discussed in the previous section, it is reasonable to require that such independent preparations must entail independent estimations of the underlying momentum fields, given information of the positions. Namely, the corresponding estimator and estimation error for the estimation of the momentum of one system, must be naturally independent of the position of the other system, prepared independently of the first. One can see that, in independent preparations of two systems, assuming that $S(q_1,q_2)$ is decomposable as in Eq. (\ref{phase decomposability for independent preparations}) and $\rho_{\tilde{p}}(q_1,q_2)$ is separable as in Eq. (\ref{separable amplitude for independent preparation}), the estimation of the momentum fields with the estimator of Eq. (\ref{weakly unbiased best estimator}) and estimation error of Eq. (\ref{estimation error}), are indeed independent, satisfying the principle of estimation independence defined above. Namely, inserting Eqs. (\ref{phase decomposability for independent preparations}) and (\ref{separable amplitude for independent preparation}) into Eqs. (\ref{weakly unbiased best estimator}) and (\ref{estimation error}), we have
\begin{eqnarray}
\overline{p}_j=\partial_{q_j}\big(S_1(q_1)+S_2(q_2)\big)=\partial_{q_j}S_j(q_j),~~~~~\nonumber\\
\epsilon_{p_j}=\frac{\xi}{2}\partial_{q_j}\ln\big(\rho_{\tilde{p}_1}(q_1)\rho_{\tilde{p}_2}(q_2)\big)=\frac{\xi}{2}\partial_{q_j}\ln\rho_{\tilde{p}_j}(q_j),
\label{satisfying estimation independence}
\end{eqnarray}
$j=1,2$; i.e., the estimator $\overline{p}_j$ and estimation error $\epsilon_{p_j}$ for the estimation of $\tilde{p}_j$ of system $j$, are indeed independent of $q_i$ of system $i$, $i\neq j$, $i,j=1,2$. Below we argue for the opposite direction, that is, assuming the principle of estimation independence of Eq. (\ref{satisfying estimation independence}) will uniquely single out the specific forms of the estimator and especially the estimation error respectively given by Eqs. (\ref{weakly unbiased best estimator}) and (\ref{estimation error}). 

Let us first apply the principle of estimation independence to single out the form the estimator. Assume initially that, given information on the positions $q$, the estimator for the momentum field takes the following general form $\overline{p}_j(q)=F_j(S(q))$, $j=1,2$, where $F_j$ is a mapping or an operator over the space of scalar functions $\Gamma$, i.e., $F_j: h(q)\in\Gamma\mapsto F_j(h(q))$, $j=1,2$. Now, we impose the principle of estimation independence which requires that, in independent preparations of two systems in which the decomposability of Eq. (\ref{phase decomposability for independent preparations}) applies, the estimator $\overline{p}_j$ for the momentum field $\tilde{p}_j$ of system $j$ is independent of the position $q_i$ of system $i$, $i\neq j$, $i,j=1,2$. This means that the estimator must fullfil the following constraint: 
\begin{eqnarray}
\overline{p}_j(q)&=&F_j\big(S(q_1,q_2)\big)\nonumber\\
&=&F_j\big(S_1(q_1)+S_2(q_2)\big)=F_j\big(S_j(q_j)\big), 
\label{principle of EI for phase}
\end{eqnarray}
$j=1,2$. Assuming that the mapping $F_j$ is linear, the above functional equation is solved by any local mapping $F_j$ satisfying $F_j(h(q_i))=\delta_{ij}F_j(h(q_i))$, $i,j=1,2$, where $\delta_{ij}$ is the Kronecker delta. Requiring importantly that $\overline{p}_j(q)=F_j(S(q))$ has a smooth classical limit recovering Eq. (\ref{HJ condition}) in the macroscopic physical regime, the most natural choice is $F_j=\partial_{q_j}$, so that we have
\begin{eqnarray}
\overline{p}_j(q)=F_j\big(S(q_1,q_2)\big)=\partial_{q_j}S(q_1,q_2), \nonumber
\end{eqnarray}
$j=1,2$, which is just the estimator postulated in Eq. (\ref{weakly unbiased best estimator}).  

Next, we show that the principle of estimation independence also singles out the specific estimation error postulated in Eq. (\ref{estimation error}), for the estimation of momentum given information on positions. We first note that the estimation error should naturally depend on the preparation setting, hence it should depend on the underlying momentum field $\tilde{p}(q;\xi)$ arising in the preparation. Due to the epistemic restriction, which asserts that the momentum field irreducibly correlates with the distribution of positions, the estimation error should therefore depend on the probability distribution of positions $\rho_{\tilde{p}}(q)$. Furthermore, it is reasonable to assume that the estimation error is independent of the estimator $\overline{p}(q)=\partial_qS(q)$. Noting this, we assume that the estimation error for estimating $\tilde{p}(q;\xi)$ with the estimator $\overline{p}(q)=\partial_qS(q)$ takes the following general form $\epsilon_p(q;\xi)=\partial_qG\big(\rho_{\tilde{p}}(q);\xi\big)$, where $G$ is some scalar function. This general form also guarantees that the estimation error transforms in the same way as that of the estimator. Our task in then to fix the functional form of $G\big(\rho_{\tilde{p}}(q);\xi\big)$ by imposing the principle of estimation independence. 

To do this, consider an ensemble of independent preparations of two systems so that the probability distribution of positions $\rho_{\tilde{p}}(q)$ is separable as in Eq. (\ref{separable amplitude for independent preparation}). The principle of estimation independence then demands that, in this independent preparations, the error $\epsilon_{p_j}$ of estimation of the momentum field $\tilde{p}_j$ of system $j$, is independent of the position $q_i$ of system $i$, $i\neq j$, $i,j=1,2$, i.e.,
\begin{eqnarray}
\partial_{q_j}G\big(\rho_{\tilde{p}}(q_1,q_2);\xi\big)&=&\partial_{q_j}G\big(\rho_{\tilde{p}_1}(q_1)\rho_{\tilde{p}_2}(q_2);\xi\big)\nonumber\\
&=&\partial_{q_j}G\big(\rho_{\tilde{p}_j}(q_j);\xi\big), 
\label{principle of EI for amplitude}
\end{eqnarray}
$j=1,2$. Integrating the above differential equation, and discarding the irrelevant constant of integration, $G$ must therefore satisfy the following relation, for any $\xi$:
\begin{equation}       
G\big(\rho_{\tilde{p}_1}(q_1)\rho_{\tilde{p}_2}(q_2);\xi\big)=G\big(\rho_{\tilde{p}_1}(q_1);\xi\big)+G\big(\rho_{\tilde{p}_2}(q_2);\xi\big).  \nonumber
\label{additivity}
\end{equation}
The above functional equation can be solved to give 
\begin{equation} 
G\big(\rho_{\tilde{p}}(q,t);\xi\big)=\gamma(\xi)\ln\rho_{\tilde{p}}(q,t),  \nonumber
\label{stochastic entropy}
\end{equation}
up to a free parameter $\gamma(\xi)$ of action dimensional which depends solely on the global random variable $\xi$. Finally, inserting back into the initial assumption $\epsilon_p(q;\xi)=\partial_qG\big(\rho_{\tilde{p}}(q);\xi\big)$, we have 
\begin{equation} 
\epsilon_p(q;\xi)=\gamma(\xi)\partial_q\ln\rho_{\tilde{p}}(q,t). 
\label{estimation error with free parameter}
\end{equation}
The estimation error postulated in Eq. (\ref{estimation error}) is a specific case of Eq. (\ref{estimation error with free parameter}) when the free parameter takes a specific form $\gamma(\xi)=\xi/2$. Moreover, in this case, to reproduce quantum uncertainty, the fluctuation of $\xi$ must meet the condition of Eq. (\ref{Planck constant}). We note that, the above strategy to single out the specific form of estimation error from the assumption of estimation independence is similar to that additivity for independent systems singles out Shannon information entropy.

Let us summarize the above result. First, to have a smooth quantum-classical correspondence, we have upgraded the intuitive basic features of classical mechanics in Hamilton-Jacobi formalism | i.e., that the Hamilton's principal function and the distribution of positions for independent preparations are respectively decomposable and separable | to somehow also apply in the microscopic world expressed in Eqs. (\ref{phase decomposability for independent preparations}) and (\ref{separable amplitude for independent preparation}). We then assume a statistical model with a fundamental epistemic restriction so that the allowed forms of distribution of positions are irreducibly parameterized by the underlying random momentum fields, and consider the estimation of the momentum field given information on the conjugate positions. We showed that, requiring a transparent and plausible principle of estimation independence which demands that independent preparations must impose independent estimations, i.e., the corresponding estimators and estimation errors are independent, single out the estimator of Eq. (\ref{weakly unbiased best estimator}), and especially the specific estimation error of Eq. (\ref{estimation error}) up to the statistics of a global-nonseparable random parameter $\gamma(\xi)$.     

It is instructive to compare the above principle of estimation independence with the principle of preparation independence underlying the PBR theorem \cite{PBR no psi-epistemic}. First, note that within the statistical model, the measurement outcome is determined by the initial position, initial momentum field, and crucially by a finite time fluctuation of $\xi(t)$, denoted below as $[\xi(t)]$ \cite{Agung-Daniel model,Agung epistemic interpretation}. They constitute the {\it instrumental} hidden variables of the model. The distribution of the instrumental hidden variables determining the measurement outcome associated with independent preparations of two systems is thus given by, noting Eqs. (\ref{phase decomposability for independent preparations}) and (\ref{separable amplitude for independent preparation}),  
\begin{eqnarray}
{\text P}_{\{S,\rho_{\tilde{p}}\}}\big(p,q,[\xi(t)]\big)&=&\prod_{i=1,2}\delta\Big(p_i-\partial_{q_i}S_i-\frac{\xi}{2}\frac{\partial_{q_i}\rho_{\tilde{p}_i}}{\rho_{\tilde{p}_i}}\Big)\nonumber\\
&\times&\rho_{\tilde{p}_i}(q_i)\chi[\xi(t)],
\label{distribution of instrumental hidden variable}
\end{eqnarray}
where $\chi[\xi(t)]$ is the probability density that $[\xi(t)]$ occurs. One can then see that, because of the nonseparability (i.e. globalness) of $\xi$, thus the nonseparability of $[\xi(t)]$, the distribution of the instrumental hidden variables associated with independent preparations is not factorizable, violating the principle of preparation independence, i.e., 
\begin{eqnarray}
&&{\text P}_{\{S,\rho_{\tilde{p}}\}}\big(p,q,[\xi(t)]\big)\nonumber\\
&\neq&{\text P}_{\{S_1,\rho_{\tilde{p}_1}\}}\big(p_1,q_1,[\xi(t)]\big){\text P}_{\{S_2,\rho_{\tilde{p}_2}\}}\big(p_2,q_2,[\xi(t)]\big).
\label{violation of preparation independence}
\end{eqnarray}

It is also clear from the above examination that the principle of preparation independence is regained when, unlike our model, $\xi$ is separable into two independent random variables, so that we have $[\xi(t)]=([\xi_1(t)],[\xi_2(t)])$ with $\chi([\xi(t)])=\chi_1([\xi_1(t)])\chi_2([\xi_2(t)])$. This is consistent with the result shown in Ref. \cite{Agung-Daniel model} that, within the epistemic reconstruction, the nonseparability of $\xi$ is indeed indispensable for describing interacting systems yielding quantum entanglement. Note further that the principle of preparation independence is also approximately recovered in the classical limit $|\partial_qS(q)|\gg|\frac{\xi}{2}\partial_q\ln\rho_{\tilde{p}}(q)|$ so that the estimation error is ignorable. In the above two cases, the probability distribution of instrumental hidden variables of Eq. (\ref{distribution of instrumental hidden variable}) becomes factorizable as required by the PBR theorem. Hence, the principle of estimation independence proposed in the present work is, in the above sense, weaker than the principle of preparation independence underlying the PBR theorem. 

Finally, let us give a simple concrete example of a scheme of estimation of momentum field with an estimation error that violates the principle of estimation independence, yielding an uncertainty relation which differs from Eq. (\ref{trade off between MS error of position and momentum}), and modifies nontrivially the Heisenberg-Kennard uncertainty relation of Eq. (\ref{HK UR}). Consider an estimation scheme for a preparation so that the estimator for the momentum field takes the form of Eq. (\ref{weakly unbiased best estimator}), but with an estimation error having a form that is different from Eq. (\ref{estimation error}), given by 
\begin{eqnarray}
\epsilon_{p_j}(q;\xi,\Lambda)&=&\tilde{p}(q;\xi)-\partial_qS(q)\nonumber\\
&=&\frac{\xi}{2}\frac{\partial_{q_j}\rho_{\tilde{p}}(q)}{\rho_{\tilde{p}}(q)}\big(1+\Lambda\rho_{\tilde{p}}(q)\big), 
\label{impossible estimation error}
\end{eqnarray}
$j=1,\dots,N$, where $\Lambda$ is some dimensionless real parameter. Comparing Eq. (\ref{impossible estimation error}) with the estimation error of Eq. (\ref{estimation error}), we have thus added a multiplicative correction term parameterized by $\Lambda$, and Eq. (\ref{estimation error}) is regained as a specific case when $\Lambda=0$. The above statistical model still has a smooth classical limit. Namely, in the physical regime when the estimation error is much smaller than the estimator, we regain the classical mechanics relation: $\tilde{p}\approx\overline{p}=\partial_qS$. Moreover, noting that $\rho_{\tilde{p}}(q)^2$ is also vanishing at the boundary, the average of the above estimation error over $\rho_{\tilde{p}}(q)$ is vanishing, hence the estimation is weakly unbiased. 

Now, consider two systems with a spatial configuration $q=(q_1,q_2)$, and suppose that they are prepared independently of each other so that the probability distribution of positions is separable as in Eq. (\ref{separable amplitude for independent preparation}). In this case, one can see that the choice of estimation error of Eq. (\ref{impossible estimation error}) does not respect the principle of estimation independence, i.e.,
\begin{eqnarray}
&&\epsilon_{p_j}\big(\rho_{\tilde{p}}(q);\xi,\Lambda\big)=\epsilon_{p_j}\big(\rho_{\tilde{p}_1}(q_1)\rho_{\tilde{p}_2}(q_2);\xi,\Lambda\big)~~~~\nonumber\\
&=&\frac{\xi}{2}\frac{\partial_{q_j}\rho_{\tilde{p}_j}(q_j)}{\rho_{\tilde{p}_j}(q_j)}\big(1+\Lambda\rho_{\tilde{p}_1}(q_1)\rho_{\tilde{p}_2}(q_2)\big)\nonumber\\
&\neq&\frac{\xi}{2}\frac{\partial_{q_j}\rho_{\tilde{p}_j}(q_j)}{\rho_{\tilde{p}_j}(q_j)}\big(1+\Lambda\rho_{\tilde{p}_j}(q_j)\big)\nonumber\\
&=&\epsilon_{p_j}\big(\rho_{\tilde{p}_j}(q_j);\xi,\Lambda\big),
\label{correlation of error without interaction}
\end{eqnarray}
$j=1,2$. Namely, even when the two systems are prepared independently of each other, the estimation error $\epsilon_{p_j}$ of estimating the momentum field $\tilde{p}_j$ of system $j$ depends on the position $q_i$ of system $i$, where $i\neq j$, $i,j=1,2$.  

Next, consider for simplicity one spatial dimension. Then, in the scheme of estimation of momentum with the estimation error of Eq. (\ref{impossible estimation error}), the associated MS error reads 
\begin{eqnarray}
\mathcal{E}_p^2=\frac{\hbar^2}{4}J_q+C, 
\label{violation of information trade-off}
\end{eqnarray}
where we have used Eq. (\ref{Planck constant}), $J_q$ is the Fisher information about the mean position defined in Eq. (\ref{Fisher information about the mean position}), and $C$ is a functional of $\rho_{\tilde{p}}(q)$ defined as $C[\rho_{\tilde{p}}(q)]\doteq\frac{\hbar^2}{4}\int{\rm d}q\,\big(\frac{\partial_q\rho_{\tilde{p}}}{\rho_{\tilde{p}}}\big)^2(2\Lambda\rho_{\tilde{p}}+\Lambda^2\rho_{\tilde{p}}^2)\rho_{\tilde{p}}$. $C$ thus gives a nontrivial correction to the information trade-off of Eq. (\ref{information trade-off}). Clearly, we have in general a nontrivial correction $C\neq 0$ for $\Lambda\neq 0$. For all $\Lambda$, $C$ is vanishing for a spatially nonuniform $\rho_{\tilde{p}}(q)$, but such a $\rho_{\tilde{p}}(q)$ is not normalized, hence is ruled out. On the other hand, as also mentioned in the previous section, in an estimation of mean position $q_{o}$ with the unbiased estimator $q$, the associated MS error must satisfy the Cram\'er-Rao inequality: $\mathcal{E}_{q}^2\doteq\int{\rm d}q(q-q_{o})^2\rho_{\tilde{p}}(q)\ge 1/J_{q}$. Combining this with Eq. (\ref{violation of information trade-off}), we thus obtain 
\begin{eqnarray}
\mathcal{E}_p^2\mathcal{E}_q^2\ge\frac{\hbar^2}{4}+\mathcal{E}_q^2C\ge\frac{\hbar^2}{4}+\frac{C}{J_q}. 
\label{modified trade-off MS errors}
\end{eqnarray}
which is quantitatively and qualitatively different from Eq. (\ref{trade off between MS error of position and momentum}). 

Now, we recall that within the estimation scheme of Sec. \ref{Quantum uncertainty as a fundamental limitation on simultaneous estimations under epistemic restriction}, wherein the model reproduces the standard quantum mechanics \cite{Agung epistemic interpretation}, we have shown in Eq. (\ref{measurement as unbiased estimates}) that the variances of momentum and position are equal to that obtained in momentum and position measurements \cite{Agung-Daniel model}. In general, it can be shown that, within the estimation scheme with the estimator and estimation error respectively given by Eqs. (\ref{weakly unbiased best estimator}) and (\ref{estimation error}), the average of physical quantity $O(p,q)$ up to second order in $p$ is equal to the average of the outcome of measurement of the associated quantum observable $\hat{O}$, i.e., $\braket{O}_{\{S,\rho_{\tilde{p}}\}}=\braket{\psi|\hat{O}|\psi}$ \cite{Agung-Daniel model}. Namely, while each single measurement outcome does not yield the objective value of $O$ prior to measurement, its average recovers the average of $O$. Let us then assume that this conclusion for the estimation scheme with the estimation error of Eq. (\ref{estimation error}) can be carried over to the modified estimation scheme with the estimation error given by Eq. (\ref{impossible estimation error}). Hence, we assume that in this modified estimation scheme, any reliable measurement of a physical quantity $O(p,q)$, up to second order in momentum, must yield outcomes which also reproduce the mean value of $O$ over the distribution of $(p,q)$ of the underlying model. 

As a corollary of the above assumption, the variances of the outcomes of momentum and position measurements must reproduce the variances of $p$ and $q$ of the statistical model denoted respectively again by $\sigma_p^2$ and $\sigma_q^2$. To study the uncertainty relations between the measurement outcomes of momentum and position, it is thus sufficient to derive the uncertainty relations between $\sigma_p^2$ and $\sigma_q^2$. First, from Eqs. (\ref{impossible estimation error}) and (\ref{Planck constant}), the variance of momentum can be computed to get $\sigma_p^2=\mathcal{E}_p^2+\Delta_p^2$, where $\Delta_p^2$ is the dispersion of the estimator $\partial_qS(q)$ defined in the previous section, and $\mathcal{E}_p^2$ is given by Eq. (\ref{violation of information trade-off}). On the other hand, one has $\sigma_q^2=\mathcal{E}_q^2$. Combining the two equations, and noting Eq. (\ref{modified trade-off MS errors}), we finally obtain
\begin{eqnarray}
\sigma_p^2\sigma_q^2\ge\frac{\hbar^2}{4}+\Delta_p^2\mathcal{E}_q^2+\frac{C}{J_q}. 
\label{modified Heisenberg-Kennard UR}
\end{eqnarray}
One can see that the last term on the right hand side of Eq. (\ref{modified Heisenberg-Kennard UR}) supplies a non-trivial correction to the Heisenberg-Kennard uncertainty relation of Eq. (\ref{HK UR}). We shall show in a different work that the modified estimation error of the kind of Eq. (\ref{impossible estimation error}) also leads to a nonlinear modification of the Schr\"odinger equation.   
            
\section{Conclusions and Discussions} 

We have worked within a general epistemic framework for a broad class of nonclassical theories by introducing a fundamental epistemic restriction to an otherwise classical theory so that the allowed forms of distribution of positions are irreducibly parameterized by the forms of the underlying momentum fields \cite{Agung-Daniel model}. Quantum mechanics is shown as a specific nonclasical theory in this epistemic framework which operationally emerges when the agent demands a specific, weakly unbiased, best estimation of the momentum field arising in her preparation, given information on the conjugate positions [by minimizing the MS error] \cite{Agung epistemic interpretation}. In particular, quantum uncertainty finds its epistemic origin from the irreducible trade-off between the MS errors of simultaneous estimation of momentum field and mean position, which in turn is directly obtained from the specific forms of estimator and estimation error for the estimation of momentum given the positions, respectively given by Eqs. (\ref{weakly unbiased best estimator}) and (\ref{estimation error}). We then argued that these specific forms of estimator and especially estimation error, can be motivated by imposing a transparent, intuitively appealing, and plausible requirement of estimation independence which demands that independent preparations must entail independent estimations. In this sense, estimation independence can be regarded as one of the physical principles implying the exact form of quantum uncertainty.  

The main result argued in the present paper suggests that a class of modifications of quantum mechanics | e.g. those implying a deviation from the exact form of quantum uncertainty | may have to violate the plausible principle of estimation independence, which is very unlikely to occur. Intuitively, like the informational axioms developed in the generalized probabilistic theories \cite{Short generalized entropy,Kimura generalized entropy,Barnum generalized entropy}, e.g. information causality \cite{Pawlowski informational approach Tsirelson bound} or data processing inequality \cite{Dahlsten DPI,Wakakuwa GMI,Al-Safi DPI}, the principle of estimation independence is also a constraint which forbids implausible information gain. Namely, a violation of estimation independence may allow an agent to benefit from the information on the position of a system, to reduce the error of estimation of the momentum of another system, prepared independently of the first. It might therefore provide a starting point to investigate why a deviation from quantum uncertainty relation may allow for the construction of a perpetual machine violating the second law of thermodynamics \cite{Hanggi UR second law}. Moreover, since a violation of quantum uncertainty may lead to stronger than quantum correlation  \cite{Oppenheim-Wehner entropic UR and QS,Ver Steeg relaxing uncertainty relation}, which in turn may imply implausible computational power \cite{Popescu review,Dam informational approach Tsirelson bound,Brassard informational approach Tsirelson bound,Buhrman superstrong cryptography,Linden nonlocal computation,Brunner trivial communication,Pawlowski informational approach Tsirelson bound,Gross trivial dynamics with superstrong correlation}, it is also interesting to see if the principle of estimation independence which uniquely defines the quantum uncertainty can be applied to also single out the specific set of quantum correlations. We note that since the principle of estimation independence implies quantum mechanics which respects no-signalling, it should be stronger than the principle of no-signalling which allows stronger than quantum correlation as exemplified by the PR boxes \cite{Popescu-Rohrlich axioms,Popescu review}. 

Within the epistemic reconstruction, it becomes clear that Planck constant arises as a free parameter whose numerical value is not fixed by the principle of estimation independence. The value of Planck constant must be determined via experiments, or by devising a deeper theory in which the Planck constant is computable. This is related to a tantalizing question: what is the physical nature of the global ontic random variable $\xi$ satisfying Eq. (\ref{Planck constant}), and why Nature, or an agent's description of It, suffers from an epistemic restriction in the first place \cite{Agung-Daniel model,Agung epistemic interpretation}? Furthermore, it is remarkable that, while Planck constant is not fixed by the principle of estimation independence, to obtain quantum mechanics within the epistemic framework, the Planck constant must indeed be constant in $(q,t)$ \cite{Agung-Daniel model}. This suggests that we may obtain a nontrivial extension of quantum mechanics, without violating the principle of estimation independence, by allowing the Planck constant fluctuates on spacetime scales much smaller than that currently observable. It is also interesting to note that the specific forms of estimator of Eq. (\ref{weakly unbiased best estimator}) and especially the estimation error of Eq. (\ref{estimation error}), are singled out in the setting of many (two or more) systems, exploiting its causal-logical informational structure. This shows that quantum mechanics is essentially a statistical operational theory for many systems, and its formalism inherits the causal-logical informational relations between the systems within the scheme of parameter estimation. 
 
Finally, Bell once speculated on the reason why Einstein did not like Bohmian mechanics \cite{Bell's on Einstein on Bohm}: ``I think Einstein thought that Bohm's model was too glib | too simple. I think he was looking for a much more profound rediscovery of quantum phenomena. The idea that you could just add a few variables and the whole thing [quantum mechanics] would remain unchanged apart from the interpretation, which was a kind of trivial addition to ordinary quantum mechanics, must have been a disappointment to him. [...]. I am sure that Einstein, and most other people, would have liked to have seen some `big principle' emerging, like the principle of `relativity', or the principle of the `conservation of energy'. In Bohm's model one did not see anything like that.'' Submitting to this spirit, we have offered an epistemic reconstruction of quantum mechanics by employing a set of physically transparent principles. Two of them are a form of causal-logical informational (inferential) principle of `estimation independence' discussed in the present work, and an invariance principle of `conservation of average energy' discussed in Refs. \cite{Agung-Daniel model,Agung epistemic interpretation}. These two principles are shared in common also by classical mechanics. The other principle, which is foreign to classical mechanics, is the assumption of epistemic restriction parameterized by a global-nonseparable random variable on the order of Planck constant \cite{Agung-Daniel model,Agung epistemic interpretation}.    

\begin{acknowledgments}

The present work is partially supported by the John Templeton Foundation (project ID 43297). The opinions expressed in this publication do not necessarily reflect the views of the John Templeton Foundation. It is also in part supported by the Indonesia Ministry of Research, Technology, and Higher Education (MRTHE) through PDUPT research scheme (Contract No. 2/E1/KP.PTNBH/2019), and the Institut Teknologi Bandung WCU Program. I would like to thank Daniel Rohrlich for continuous support and insightful and stimulating advice, and also to Joseph Emerson, Michael Hall, Mile Gu, Guy Hetzroni, Varun Narasimhachar, and Hermawan K. Dipojono for helpful discussions on various versions and aspects of the present work. 

\end{acknowledgments}

\appendix

\section{Schr\"odinger-Robertson uncertainty relation between position and momentum\label{Schroedinger-Robertson uncertainty relation}}

Within the epistemic reconstruction based on the estimation scheme with the estimator and estimation error given respectively by Eqs. (\ref{weakly unbiased best estimator}) and (\ref{estimation error}), the Schr\"odinger-Robertson uncertainty relation for position and momentum quantum observables can be obtained as follows. For simplicity, we consider a system of one spatial dimension. First, using the language of Hilbert space formalism, the MS error for estimation of momentum can be written as $\mathcal{E}_{p}^2=\frac{\hbar^2}{4}\int{\rm d}q\big(\partial_{q}\ln\rho_{\tilde{p}}\big)^2\rho_{\tilde{p}}(q)=\int {\rm d}q\Big(\frac{\braket{\psi|[\hat{\pi}_q,\hat{p}]|\psi}}{2i|\braket{q|\psi}|^2}\Big)^2|\braket{q|\psi}|^2$, where $\hat{\pi}_q\doteq\ket{q}\bra{q}$, $[\hat{\pi}_q,\hat{p}]\doteq \hat{\pi}_q\hat{p}-\hat{p}\hat{\pi}_q$ is the commutator, and we have used Eq. (\ref{wave function}). On the other hand, the MS error for the estimation of mean position $q_{o}$ with the estimator $q$ reads $\mathcal{E}_{q}^2\doteq\int{\rm d}q(q-q_{o})^2\rho_{\tilde{p}}(q)=\int{\rm d}q(q-q_{o})^2|\braket{q|\psi}|^2$. Combining the two and using the Cauchy-Schwarz inequality, one obtains the Robertson-like uncertainty relation \cite{Robertson UR}: 
\begin{eqnarray}
\mathcal{E}_{p}^2\mathcal{E}_{q}^2\ge\frac{1}{4}\big|\braket{\psi|[\hat{q},\hat{p}]|\psi}\big|^2. 
\label{Robertson-like UR}
\end{eqnarray}
On the other hand, using again Cauchy-Schwarz inequality, one has
\begin{eqnarray}
\Delta_{p}^2\mathcal{E}_{q}^2&\doteq&\int{\rm d}q\Big(\partial_{q}S-\int{\rm d}q'\partial_{q'}S\rho_{\tilde{p}}(q')\Big)^2\rho_{\tilde{p}}(q)\nonumber\\
&\times&\int{\rm d}q\Big(q-\int{\rm d}q'q'\rho_{\tilde{p}}(q')\Big)^2\rho_{\tilde{p}}(q)\nonumber\\
&\ge&\Big|\int{\rm d}q\Big(\partial_{q}S-\int{\rm d}q'\partial_{q'}S\rho_{\tilde{p}}(q')\Big)\nonumber\\
&\times&\Big(q-\int{\rm d}q'q'\rho_{\tilde{p}}(q')\Big)\rho_{\tilde{p}}(q)\Big|^2\nonumber\\
&=&\Big|\int{\rm d}q\big(q\partial_{q}S\big)\rho_{\tilde{p}}(q)-\int{\rm d}qq\rho_{\tilde{p}}(q)\int{\rm d}q\partial_qS\rho_{\tilde{p}}(q)\Big|^2\nonumber\\
&=&\Big|\frac{1}{2}\braket{\psi|\{\hat{q},\hat{p}\}|\psi}-\braket{\psi|\hat{q}|\psi}\braket{\psi|\hat{p}|\psi}\Big|^2, 
\label{Schroedinger correction term}
\end{eqnarray}
where we have again used Eq. (\ref{wave function}), and $\{\hat{q},\hat{p}\}\doteq \hat{q}\hat{p}+\hat{p}\hat{q}$ is the anticommutator. Combining Eqs. (\ref{Robertson-like UR}) and (\ref{Schroedinger correction term}), we thus obtain the Schr\"odinger uncertainty relation \cite{Schroedinger UR}: 
\begin{eqnarray}
&&\sigma_{p}^2\sigma_{q}^2=\sigma_{\hat{p}}^2\sigma_{\hat{q}}^2=\mathcal{E}_{p}^2\mathcal{E}_{q}^2+\Delta_{p}^2\mathcal{E}_{q}^2\nonumber\\
&\ge&\frac{1}{4}\big|\braket{\psi|[\hat{q},\hat{p}]|\psi}\big|^2+\big|\frac{1}{2}\braket{\psi|\{\hat{q},\hat{p}\}|\psi}-\braket{\psi|\hat{q}|\psi}\braket{\psi|\hat{p}|\psi}\big|^2.\nonumber\\
\end{eqnarray}

\end{document}